\documentclass{iau}
\usepackage{graphicx}

\title[A mid-infrared exploration of active galactic nuclei] 
{A mid-infrared exploration of the dusty environments of active
  galactic nuclei}

\author[Almudena Alonso-Herrero]   
{Almudena Alonso-Herrero$^{1,2}$}

\affiliation{$^1$Instituto de F\'{\i}sica de Cantabria, CSIC-UC, 39005
  Santander, Spain\\ email: {\tt aalonso@ifca.unican.es} \\[\affilskip]
$^2$Augusto G. Linares Senior Research Fellow}

\pubyear{2014}
\volume{xxx}  
\pagerange{119--126}
\jname{IAU 304: Multiwavelegth AGN Surveys and Studies}
\editors{A.C. Editor, B.D. Editor \& C.E. Editor, eds.}
\begin{document}

\maketitle

\begin{abstract}
We present the first results from a mid-infrared survey of local
Active Galactic Nuclei (AGN) using the CanariCam (CC) instrument on
the 10.4\,m Gran Telescopio 
Canarias (GTC). We are obtaining sub-arcsecond angular
resolution ($0.3-0.6\,$arcsec) mid-IR imaging and spectroscopic
observations of a sample of 100 local AGN, which are  complemented
with data taken with T-ReCS, VISIR, and 
Michelle. The full sample contains approximately 140 AGN, covers nearly six
orders of magnitude in AGN luminosity, and 
includes low-luminosity AGN (LLAGN), Seyfert 1s and 2s, QSO, radio
galaxies, and (U)LIRGs. The main goals of this 
project are:   (1) to test whether the properties of the dusty tori of
the AGN Unified Model  depend on the AGN type, (2) to study the
nuclear star formation activity and obscuration of local AGN, and 
(3) to explore the role of the dusty torus in LLAGN. 

\end{abstract}

\firstsection 
\section{Introduction}

In the context of the Unified Model \cite[(Antonucci
1993)]{Antonucci1993} for active galactic nuclei (AGN), it is
generally accepted that the obscuring 
material is located in an optically thick dusty torus. From the
modelling of nuclear infrared 
(IR) observations it has become clear that AGN tori are 
relatively compact (from a few parsecs to a few tens of
parsecs). Moreover, the torus models reproduce better the observations
if the dust is 
mostly distributed in clumps rather 
than homogeneously \cite[(see e.g., Tristram et al. 2009, Ramos Almeida et
al. 2009, 2011, Alonso-Herrero et al. 2011)]{Tristram2009, RamosAlmeida2009,
  RamosAlmeida2011,
  AlonsoHerrero2011}. However, it is still an open question
 whether the properties of
the torus vary with AGN type and/or the luminosity of the 
AGN. In addition, the tori are not isolated structures in the
nuclear regions of AGN. Rather, IR observations have shown that 
there is some continuity between the dust in the torus and that in the
nuclear environments \cite[(Roche et al. 2006, Packham et
al. 2007)]{Roche2006,Packham2007}. Finally, models
 also predict that star formation (SF) should occur inside or
near the outer edge of the dusty tori \cite[(see e.g., Wada \& Norman
2002)]{Wada2002}. 

The dust in the torus absorbs a large fraction of the AGN emission and
re-emits it in the IR peaking at mid-IR ($\sim 8-40\,\mu$m)
wavelengths. Therefore, sub-arcsecond mid-IR observations 
offer the best opportunity to constrain the torus models and learn
about the dusty environments of AGN.  The angular resolution achieved
with  ground-based 8-10\,m class telescopes, although it is not
sufficiently high to spatially resolve the torus structure, can help
 disentangle the  torus emission from the 
circumnuclear and host galaxy contribution \cite[(see e.g.,  Packham et
al. 2005, Roche et al. 2006)]{Packham2005,Roche2006}. 

We present results from a mid-IR survey of a sample of 100 local AGN using
the CanariCam \cite[(CC, Telesco et al. 2003)]{Telesco2003} instrument 
on the 10.4\,m Gran Telescopio Canarias (GTC). 
The main goal of this project is to test whether the properties of the
torus depend on the AGN type and luminosity and explore the
environments of the dusty torus by studying the nuclear star formation
(SF) and extinction produced by the host galaxy\footnote{Additional
  information about this project can be found at:  
https://sites.google.com/site/piratasrelatedpublications/}.

\section{High angular resolution ground-based mid-IR observations}
\label{sec:survey}
\begin{figure}
\begin{center}
 \includegraphics[width=4cm,angle=-90]{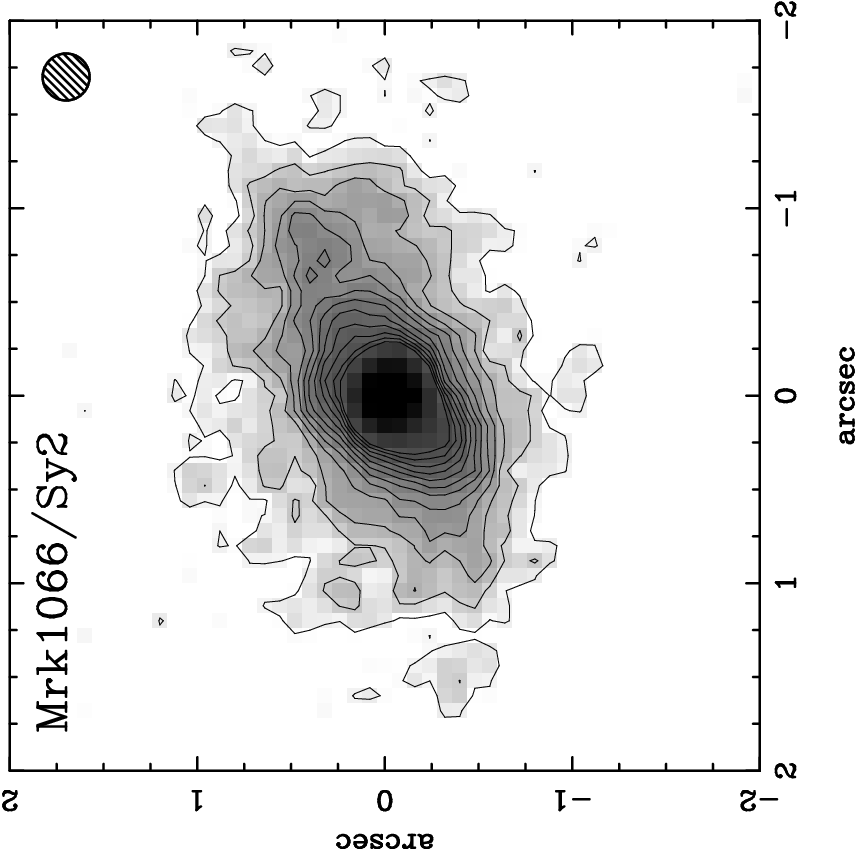} 
 \includegraphics[width=4cm,angle=-90]{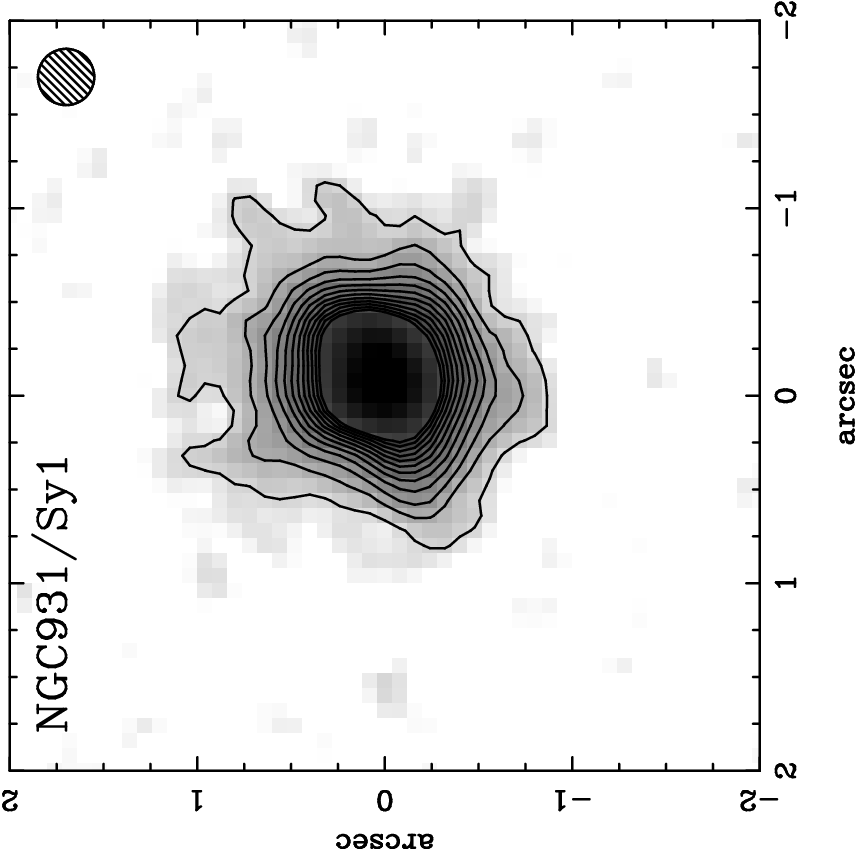} 
 \includegraphics[width=4cm,angle=-90]{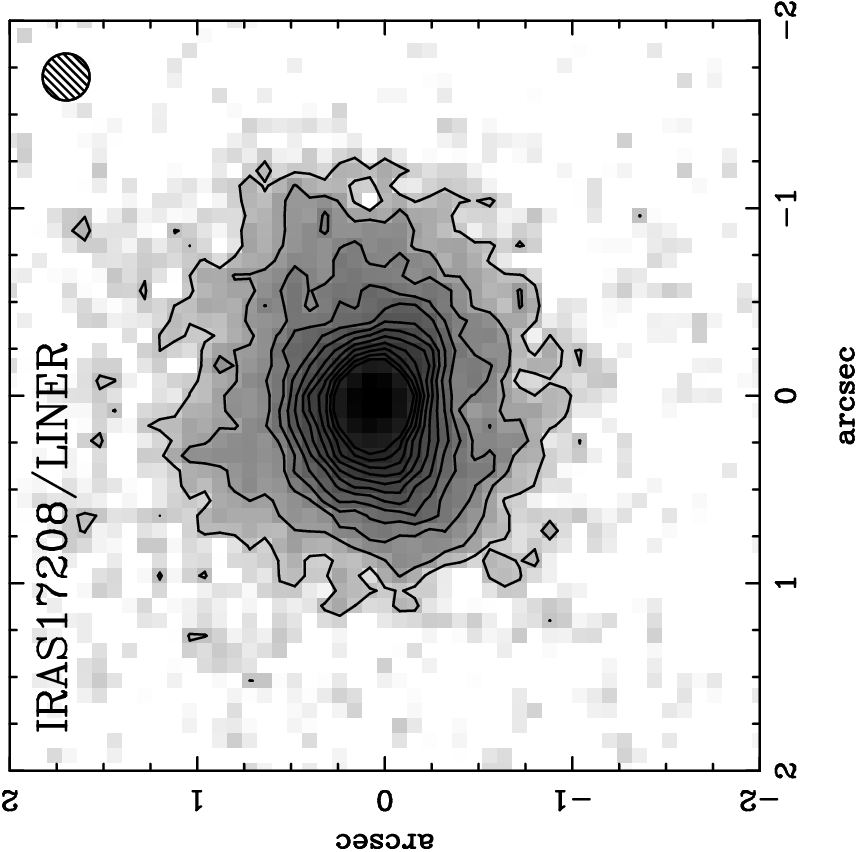} 
 \caption{GTC/CC  $8.7\,\mu$m images of three AGN in our sample. The
   hatched circles at the top right of the panels are the resolution
   of the images (FWHM). } 
   \label{fig1}
\end{center}
\end{figure}

The observations are part of the CC AGN guaranteed time 
and from an ESO/GTC large programme
(182.B-2005) for a total of ~280 hours. The observations include
high-angular resolution ($0.3-0.6\,$arcsec) imaging 
with the Si-2 ($\lambda_{\rm c}=8.7\,\mu$m) filter and 
low-resolution (nominal $R=\lambda/\Delta \lambda \simeq 175$)
spectroscopy covering the $\sim  
8-13\,\mu$m spectral range. We
complemented the CC data with existing observations with the T-ReCS,
Michelle, and VISIR instruments for approximately 40 Seyfert and
LLAGN. The full sample covers more than six decades of
AGN luminosity and includes LLAGN,
Seyfert 1s and 2s, radio galaxies, quasars and IR 
luminous and ultraluminous IR galaxies (LIRGs and ULIRGs) hosting
AGN. Fig.~\ref{fig1} shows the GTC/CC $8.7\,\mu$m images of three
AGN in our sample reduced with the CC pipeline {\sc redcan}
\cite[(Gonz\'alez-Mart\'{\i}n et al. 2013)]{GonzalezMartin2013}.

\section{Results}
\label{sec:goals}
\subsection{Torus properties of Seyfert galaxies}

Modelling the nuclear IR emission of nearby AGN with torus models can
be used  
to derive torus properties such as its geometry, distribution and
optical depth of the dust in the torus, and the AGN luminosity and covering
factor. If this is done for
a large sample of AGN, it is then possible to investigate whether such
properties depend on the AGN type (Type 1 vs. Type 2), AGN class
(AGN in LIRGs and ULIRGs, radio galaxies, Seyferts), and AGN
luminosity. Over the past few years we fitted
 sub-arcsecond angular resolution IR spectral energy distributions
(SEDs) and mid-IR spectroscopy of relatively small samples of Seyferts
and individual objects 
\cite[(see e.g., Ramos Almeida et al. 2009, 2011, Alonso-Herrero et
al. 2011,
2013)]{RamosAlmeida2009,RamosAlmeida2011,AlonsoHerrero2011,AlonsoHerrero2013}
using the \cite{Nenkova2008} 
clumpy torus models (also known as {\sc clumpy}). To deal with the
intrinsic degeneracies of  the {\sc clumpy} models, we  used the
Bayesian fitting tool
 {\sc bayesclumpy}  \cite[(Asensio Ramos \& Ramos Almeida
2009)]{AsensioRamos2009}, which allows us to derive probability
distributions of the fitted parameters (e.g., torus extent and angular
size, number and  distribution of the clouds).

\begin{figure}
\begin{center}
 \includegraphics[width=4.5cm,angle=-90]{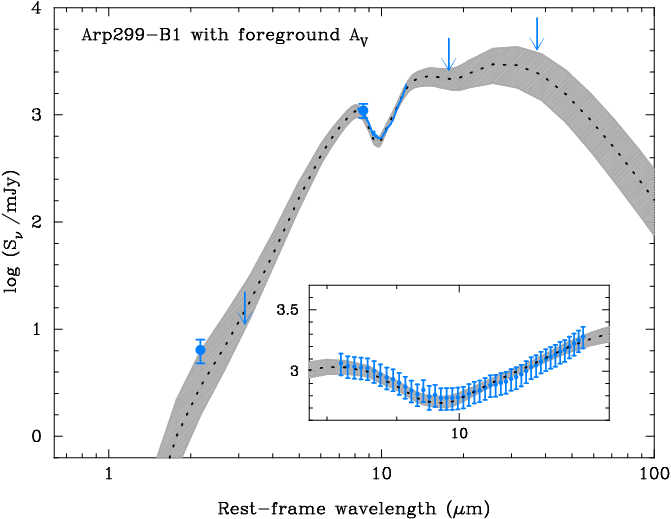} 
\hspace{0.3cm}
 \includegraphics[width=4.5cm,angle=-90]{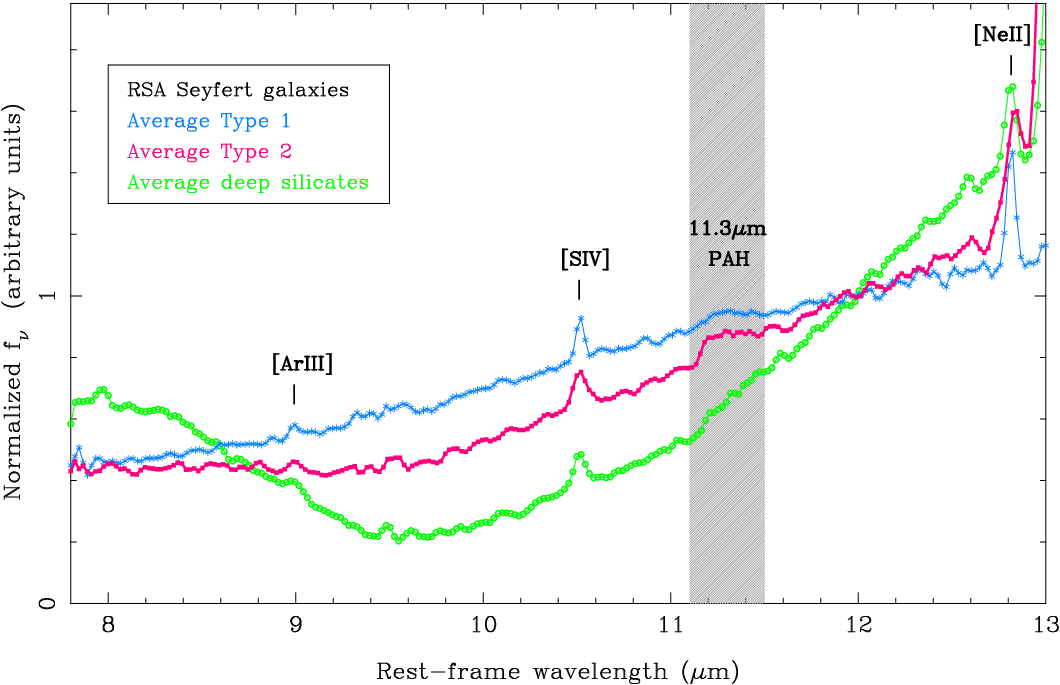} 
 \caption{{\it Left panel.} Fit to the IR SED (filled symbols and
   upper limits) and GTC/CC nuclear
   spectrum (solid line) of NGC~3690/Arp~299-B1 using the {\sc clumpy} 
   models (dotted line and shaded area) of Nenkova et al. (2008). Adapted from
   \cite{AlonsoHerrero2013}.  {\it Right panel.} 
Average nuclear spectra normalized at $12\,\mu$m 
of the Seyfert 1s (asterisks), Seyfert 2s (filled squares), 
and Seyfert nuclei with deep $10\,\mu$m silicate features (open dots)
in the RSA sample studied by \cite{Esquej2013}. We mark the most
important spectral features including the  $11.3\,\mu$m PAH feature
that is used to probe SF on nuclear scales (see Section~3.3).}  
   \label{fig2}
\end{center}
\end{figure}

Fig.~\ref{fig2} (left) shows the best fit to the IR 
SED and the GTC/CC mid-IR spectrum of the nuclear region of NGC~3690,
one of the components of the interacting LIRG Arp~299. The high
angular resolution (0.3\,arcsec) of the CC data allowed us to probe nuclear
physical scales of approximately 60\,pc, which is a factor of 10
improvement over previous mid-IR spectroscopy of this
system. From the fit done using the \cite{Nenkova2008} torus models we
inferred that the IR emission of this nucleus comes from  
 AGN-heated dust in a clumpy torus with both a high covering factor
 and high extinction along the line of sight 
 \cite[(see
  Alonso-Herrero et al. 2013 for more details)]{AlonsoHerrero2013]}.

Although so far we  studied the torus properties for relatively
small AGN samples, our recent work indicates some dependency of the 
torus geometry, in particular the covering factor, on the Type1/Type2 class
\cite[(Ramos Almeida et al. 2011)]{RamosAlmeida2011} 
 and/or the luminosity of the AGN \cite[(Alonso-Herrero et
 al. 2011)]{AlonsoHerrero2011}.  These works also highlighted the
 importance of disentangling the dust heated
 by the AGN from foreground dust in the host galaxy 
when modelling the nuclear IR emission of local AGN
 \cite[(see also, Gonz\'alez-Mart\'{\i}n et al. 2013)]{GonzalezMartin2013}.

\subsection{Nuclear Star Formation Activity in Seyfert galaxies}
Since material must 
be driven inwards from the interstellar medium of the host galaxy to
the nucleus ($<1\,$pc 
regions) to fuel the AGN, nuclear SF on scales of less than $\sim 100\,$pc    
appears to be an inevitable consequence of this process, as predicted 
by numerical simulation \cite[(see e.g., Hopkins \& Quataert
2010)]{Hopkins2010}. We can use the mid-IR polycyclic aromatic hydrocarbon
(PAH) features to probe the nuclear SF activity in local
AGN. In  particular, it has been shown that the $11.3\,\mu$m PAH
feature is not suppressed due to the presence of an AGN on kiloparsec scales
\cite[(Diamond-Stanic \& Rieke 2010)]{DiamondStanic2010}  down to
distances from the AGN of up to a few tens of parsecs \cite[(Esquej et
al. 2013)]{Esquej2013}.  

\cite{Esquej2013} compiled mid-IR nuclear spectra  obtained with 
T-ReCS, Michelle, and VISIR at sub-arcsecond angular resolutions 
for 29 Seyfert galaxies in the revised
Shapley-Ames (RSA) Seyfert sample 
\cite[(Maiolino \& Rieke 1995)]{Maiolino1995}.
Fig.~\ref{fig2} (right) shows the average
nuclear (typically probing regions of 60\,pc in size) 
spectra for the Seyfert 1s and Seyfert 2s galaxies.  Clearly, the
$11.3\,\mu$m PAH feature is detected 
in the average spectrum of both types. We also show the average
spectrum of Seyfert galaxies in the RSA sample with deep $10\,\mu$m
silicate features. Such deep features are likely due to dust in the 
host galaxy and not dust in the torus \cite[(Levenson et al. 2007,
Alonso-Herrero et  
  al. 2011, Goulding et al. 2012, Gonz\'alez-Mart\'{\i}n et
  al. 2013)]{Levenson2007, AlonsonHerrero2011, Goulding2012,
    GonzalezMartin2013}.  Nuclear $11.3\,\mu$m PAH emission  is
  detected in half
of the sample with no evidence that the emission is suppressed for more
luminous AGN.
\cite{Esquej2013} also found a relation between the nuclear SF
 rate and the black hole accretion rate (see the contribution by P. 
Esquej in these
 proceedings)  that is reproduced by
the numerical simulations of \cite{Hopkins2010} for the appropriate
physical scales probed by our data. A similar relation is observed
when the SF rate is measured on kiloparsec scales
\cite[(Diamond-Stanic \& Rieke 2012)]{DiamondStanic2012}.

\subsection{LLAGN and the origin of the torus} 
LLAGN are common in local spirals and ellipticals  \cite[(Ho
2008)]{Ho2008}. 
The detection of broad H$\alpha$ components proves
that at least some of them do contain an AGN. Interestingly, at low
luminosities ($L_{\rm bol} < 10^{42}\,{\rm erg\,s}^{-1}$) theoretical models
predict that the torus might disappear because accretion can no longer
sustain the outflow  required for large obscuring columns in the torus
\cite[(Elitzur \& Shlosman 2006)]{Elitzur2006}. 
Moreover, how the properties of LLAGN tori, if they exist, might 
relate to those in higher luminosity AGN is still not known.

Using sub-arcsecond angular resolution images \cite{Mason2012}
found a variety of mid-IR morphologies in a
sample of local LLAGN. Sources with higher
Eddington ratios tend to show more compact and bright mid-IR sources, whereas
LLAGN with low Eddington ratios appear more diffuse and
extended in the mid-IR.  It is
indeed in the former type of objects in which the torus is expected to be
detectable in our CC survey. Furthermore, from the modelling of
multi-wavelength SEDs recently
 \cite{Mason2013} 
suggested that dust may account for the nuclear mid-IR emission of
many LLAGN. We refer the reader to the contribution by R. Mason in
these proceedings for more details.

Based on observations made with the GTC, instaled in the Spanish
Observatorio del Roque de los Muchachos of the Instituto de
Astrofísica de Canarias, in the island of La Palma. 
A.A.-H. acknowledges support from the Spanish Plan Nacional through
grant AYA2009-05705-E and from the Universidad de Cantabria.

\end{document}